\documentclass[useAMS,usenatbib]{mn2e}
\usepackage{epsfig}
\usepackage{amsmath}

\newcommand{\be}{\begin{equation}}
\newcommand{\beq}{\begin{equation}}
\newcommand{\ba}{\begin{eqnarray}}
\newcommand{\ee}{\end{equation}}
\newcommand{\eeq}{\end{equation}}
\newcommand{\ea}{\end{eqnarray}}

\newcommand{\apj}{ApJ}
\newcommand{\apjl}{ApJL}
\newcommand{\mnras}{MNRAS}
\newcommand{\aj}{AJ}

\def\lsim{~\rlap{$<$}{\lower 1.0ex\hbox{$\sim$}}}

\def\gsim{~\rlap{$>$}{\lower 1.0ex\hbox{$\sim$}}}

\title[Non-Gaussianity in redshifted 21cm fluctuations]{Biased Reionisation and Non-Gaussianity in Redshifted 21cm Intensity Maps of the Reionisation Epoch}

\author[Wyithe \& Morales]{J. Stuart B. Wyithe$^1$ and Miguel F. Morales$^2$
\\$^1$ School of Physics, University of Melbourne, Parkville, Victoria,
Australia
\\$^2$ MIT Kavli Institute,  77 Massachusetts Ave.,
Cambridge, MA 02139
\\Email: swyithe@physics.unimelb.edu.au
, mmorales@space.mit.edu
}

\begin{document}


\maketitle

\label{firstpage}
\begin{abstract}

Spatial dependence in the statistics of redshifted 21cm fluctuations
promises to provide the most powerful probe of the reionisation
epoch. In this paper we consider the second and third moments of the
redshifted 21cm intensity distribution using a simple model that
accounts for galaxy bias during the reionisation process. We
demonstrate that skewness in redshifted 21cm maps should be
substantial throughout the reionisation epoch and on all angular
scales, owing to the effects of galaxy bias which leads to early
reionisation in over-dense regions of the IGM. The variance (or
power-spectrum) of 21cm fluctuations will exhibit a minimum in
redshift part way through the reionisation process, when the global
ionisation fraction is around 50\%. This minimum is generic, and is
due to the transition from 21cm intensity being dominated by
over-dense too under-dense regions as reionisation progresses. We show
that the details of the reionisation history, including the presence
of radiative feedback are encoded in the evolution of the
auto-correlation and skewness functions with redshift and mean IGM
neutral fraction. The amplitudes of fluctuations are particularly
sensitive to the masses of ionising sources, and vary by an order of
magnitude for astrophysically plausible models. We discuss the
detection of skewness by first generation instruments, and conclude
that the Mileura Widefield Array--Low Frequency Demonstrator will
have sufficient sensitivity to detect skewness on a range of angular
scales at redshifts near the end of reionisation, while a subsequent
instrument of 10 times the collecting area could map out the evolution
of skewness in detail. The observation of a minimum in variance during
the reionisation history, and the detection of skewness would both
provide important confirmation of the cosmological origin of
redshifted 21cm intensity fluctuations.
  
\end{abstract}


\section{Introduction}

The primordial density field as revealed in maps of the Cosmic
Microwave Background may be described as a Gaussian random field. In
other words, the distribution of over-densities in the Fourier
decomposition of the field is Gaussian at all spatial frequencies, so
that the statistics of the field may be described by a single number
(the variance) as a function of spatial scale. This scale dependent
variance is referred to as the power-spectrum. On large enough scales
we might also expect the statistics for 21cm emission to be Gaussian,
since over-dense regions have more hydrogen, and therefore a larger
21cm intensity. On small scales this Gaussianity will be broken by the
presence of HII regions. Thus much attention has focused on the
prospects of measuring the power-spectrum of 21cm emission, as well as
the use of the power-spectrum to probe the astrophysics of
reionisation (e.g. Zaldariaga et al.~2004; Morales \& Hewitt 2004; Furlanetto et al.~2004).

In regions that are over-dense, galaxies will be over-abundant for two
reasons: first because there is more material per unit volume to make
galaxies, and second because small-scale fluctuations need to be of
lower amplitude to form a galaxy when embedded in a larger-scale
over-density (the so-called {\it galaxy bias}; see Mo \&
White~1996). Regarding reionisation of the intergalactic medium (IGM),
the first effect will result in a larger density of ionising
sources. However this larger density will be compensated by the
increased density of gas to be ionised. Furthermore, the increase in
the recombination rate in over-dense regions will be counteracted by
the galaxy bias in over-dense regions. The process of reionisation
also contains several layers of feedback.  Radiative feedback heats
the IGM and results in the suppression of low-mass galaxy formation
(Efstathiou, 1992; Thoul \& Weinberg~1996; Quinn et al.~1996; Dijkstra
et al.~2004). This delays the completion of reionisation by lowering
the local star formation rate, but the effect is counteracted in
over-dense regions by the biased formation of massive galaxies. The
radiation feedback may therefore be more important in low-density
regions where small galaxies contribute more significantly to the
ionising flux.

Wyithe \& Loeb~(2007) have modeled the density dependent reionisation
process using a semi-analytic model that incorporates the features
described above, and so captures the important physical
processes. This model demonstrates that galaxy bias leads to enhanced
reionisation in over-dense regions. Because this bias operates on the
exponential tail of the Press-Schechter~(1974) mass function, the
enhancement of ionisation is not linear with over-density. As a
result, galaxy bias leads to a non-Gaussian distribution of 21cm
brightness temperature in intensity maps of reionisation. This effect
was discussed by Lidz et al.~(2006) using detailed numerical
simulations. Lidz et al.~(2006) found that the contribution of a third
order term in the power-spectrum will be comparable to the second
order term at an epoch where the IGM is approximately 50\% ionised.
While comparison with data may require detailed numerical simulations
of the sort described in Lidz et al.~(2006), computationally cheap
semi-analytic models are useful for exploring which physical
parameters will be probed by the detection of angular fluctuations in
the 21cm intensity. In this paper we explore the dependence of
observables on the values of parameters of interest like the clumping
of the IGM and the minimum mass for galaxy formation. We also discuss
the prospects for detection of the third moment (skewness) with first
and second generation low-frequency arrays.

In \S~\ref{model} and \S~\ref{moment} we describe our model for
density dependent reionisation, and the resulting redshifted 21cm
intensity probability distribution (including calculation of its
second and third moments). We describe the features of the
auto-correlations and skewness functions that correspond to different
physical parameters that are important for the reionisation epoch. We
then discuss the measurement of these moments using first and second
generation low-frequency arrays (in particular the Mileura Widefield
Array--Low Frequency Demonstrator) in \S~\ref{measure} before
concluding in \S~\ref{conclusion}. Throughout the paper we adopt the
set of cosmological parameters determined by {\it WMAP} (Spergel et
al. 2006) for a flat $\Lambda$CDM universe.

\section{Density Dependent model of reionisation}
\label{model}

In this paper we compute the relation between the local over-density
and the brightness temperature of redshifted 21cm emission using the
model described in Wyithe \& Loeb~(2007). Here we summarise the main features of the model and refer the reader to that paper for more details.

The evolution of the ionisation fraction by mass $Q_{\delta,R}$ of a
particular region of scale $R$ with over-density $\delta$ (at observed
redshift $z_{\rm obs}$) may be written as
\begin{eqnarray}
\label{history}
\nonumber
\frac{dQ_{\delta,R}}{dt} &=& \frac{N_{\rm ion}}{0.76}\left[Q_{\delta,R} \frac{dF_{\rm col}(\delta,R,z,M_{\rm ion})}{dt} \right.\\
\nonumber
&&\hspace{5mm}+ \left.\left(1-Q_{\delta,R}\right)\frac{dF_{\rm col}(\delta,R,z,M_{\rm min})}{dt}\right]\\
&-&\alpha_{\rm B}Cn_{\rm H}^0\left(1+\delta\frac{D(z)}{D(z_{\rm obs})}\right) \left(1+z\right)^3Q_{\delta,R},
\end{eqnarray}
where $N_{\rm ion}$ is the number of photons entering the IGM per
baryon in galaxies, $\alpha_{\rm B}$ is the case-B recombination
co-efficient, $C$ is the clumping factor (which we assume, for
simplicity, to be constant), and $D(z)$ is the growth factor between
redshift $z$ and the present time. The production rate of ionising
photons in neutral regions is assumed to be proportional to the
collapsed fraction $F_{\rm col}$ of mass in halos above the minimum
thresholds in neutral ($M_{\rm min}$), and in ionised ($M_{\rm ion}$)
regions. For our fiducial model, we assume $M_{\rm min}$ to correspond to
a virial temperature of $10^4$K, representing the hydrogen cooling
threshold, and $M_{\rm ion}$ to correspond to a virial temperature of
$10^5$K, representing the mass below which infall is suppressed from
an ionised IGM (Dijkstra et al.~2004). In a region of co-moving radius
$R$ and mean over-density $\delta(z)=\delta D(z)/D(z_{\rm obs})$
[specified at redshift $z$ instead of the usual $z=0$], the relevant
collapsed fraction is obtained from the extended
Press-Schechter~(1974) model (Bond et al.~1991) as
\begin{equation}
F_{\rm col}(\delta,R,z) = \mbox{erfc}{\left(\frac{\delta_{\rm
c}-\delta(z)}{\sqrt{2\left(\left[\sigma_{\rm
gal}\right]^2-\left[\sigma(R)\right]^2\right)}}\right)},
\end{equation}
where $\mbox{erfc}(x)$ is the error function, $\sigma(R)$ is the
variance of the density field smoothed on a scale $R$, and
$\sigma_{\rm gal}$ is the variance of the density field smoothed on a
scale $R_{\rm gal}$, corresponding to a mass scale of $M_{\rm min}$ or
$M_{\rm ion}$ (both evaluated at redshift $z$ rather than at $z=0$).
In this expression, the critical linear over-density for the collapse
of a spherical top-hat density perturbation is $\delta_c\approx 1.69$.

The model assumes that on large (linear-regime) scales most ionising
photons are absorbed locally, so that the ionisation of a region is caused
by nearby ionisation sources. This assumption is certainly justified during
the early stages of reionisation, when the mean free path for ionising
photons is short.  However even later in the reionisation process, the mean
free-path always remains smaller than the characteristic HII bubble size
(it could be smaller if mini-halos or pockets of residual HI block ionising
photons between the sources and the edge of the HII region.) Our local
ionisation assumption is therefore valid as long as the characteristic
bubble size is smaller than the spatial scale of the correlations we
consider. In this paper we consider scales of 1-25 arc-minutes which will be
of interest to upcoming 21cm experiments, corresponding to co-moving scales
of 2-50 Mpc. These scales are larger than the mean-free-path of ionising
photons near the end of reionisation (Fan et al.~2006). 

As an example, we find the value of $N_{\rm ion}$ that yields overlap
of ionised regions at the mean density IGM by $z\sim6$ (White et
al.~2003). Equation~(\ref{history}) may be integrated as a function of
$\delta$.  At a specified redshift, this yields the filling fraction
of ionised regions within the IGM on various scales $R$ as a function
of over-density. We may then calculate the corresponding 21cm
brightness temperature contrast
\begin{equation}
T(\delta,R) = 22\mbox{mK}(1-Q_{\delta,R})\left(1+\frac{4}{3}\delta\right),
\end{equation}
where the pre-factor of 4/3 on the over-density refers to the
spherically averaged enhancement of the brightness temperature due to
peculiar velocities in over-dense regions (Bharadwaj \& Ali~2005;
Barkana \& Loeb~2005).

\begin{figure}
\includegraphics[width=8.cm]{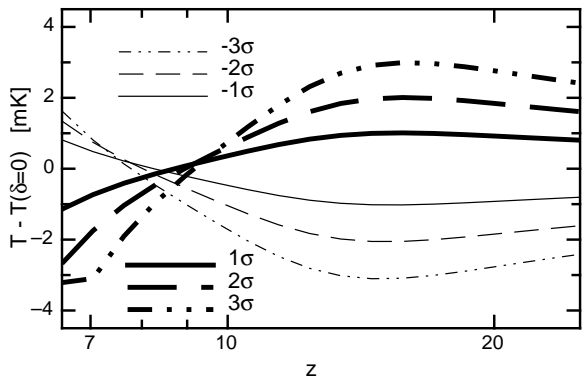} 
\caption{The difference in brightness temperature between over or under-dense regions, and the IGM at mean density as a function of redshift. The lines represent regions at redshift $z$ that are $-3\sigma$, $-2\sigma$,
$-1\sigma$, $0\sigma$, $1\sigma$, $2\sigma$ or 3$\sigma$ fluctuations
in the smoothed density field. We assumed a clumping factor of $C=10$ and an angular scale of $10^\prime$.}
\label{fig1}
\end{figure} 

The predicted evolution of the brightness temperature of regions of
IGM at fixed angular scale is plotted in Figure~\ref{fig1}. For
clarity we have plotted $T-T(\delta=0)$, i.e. the difference in
brightness temperature between the over or under-dense region, and the
IGM at mean density. Each line in the figure represents a region that
is a $-3\sigma$, $-2\sigma$, $-1\sigma$, $0\sigma$, $1\sigma$,
$2\sigma$ or 3$\sigma$ fluctuation in the smoothed density field at
redshift $z$. Here we assumed a clumping factor of $C=10$ and an
angular scale of $10^\prime$, where $\theta=R/d_{\rm A}(z)$ and
$d_{\rm A}$ is the angular diameter distance. At high redshift the
curves are symmetric about zero, as expected for linear fluctuations
in the primordial density field. As reionisation proceeds the
brightness temperature (relative to the mean IGM) is initially quite
insensitive to redshift. However at later times the brightness
temperature of over-dense regions decreases, while the brightness
temperature of under-dense regions increases (again when compared to
the mean IGM). Finally, reionisation is completed in over-dense
regions of the IGM first, as may be seen in the 3$\sigma$ fluctuation
which is reionised at $z\sim7$, prior to the mean IGM at
$z=6$. Therefore at some point during the reionisation history there
must be a transition between bright patches of 21cm emission being
observed in over-dense regions, too bright patches being observed in
under-dense regions. This transition results in an epoch of nearly
uniform brightness temperature (uniform when smoothed on an angular
scale larger than the typical HII region) of the
IGM. Figure~\ref{fig1} shows that in our models this transition occurs
near a redshift of $z\sim8-9$. The epoch of near uniform temperature
across the IGM has interesting consequences for the variance in
smoothed 21cm intensity maps. This topic will be discussed in the
following section.

\begin{figure*}
\includegraphics[width=15.cm]{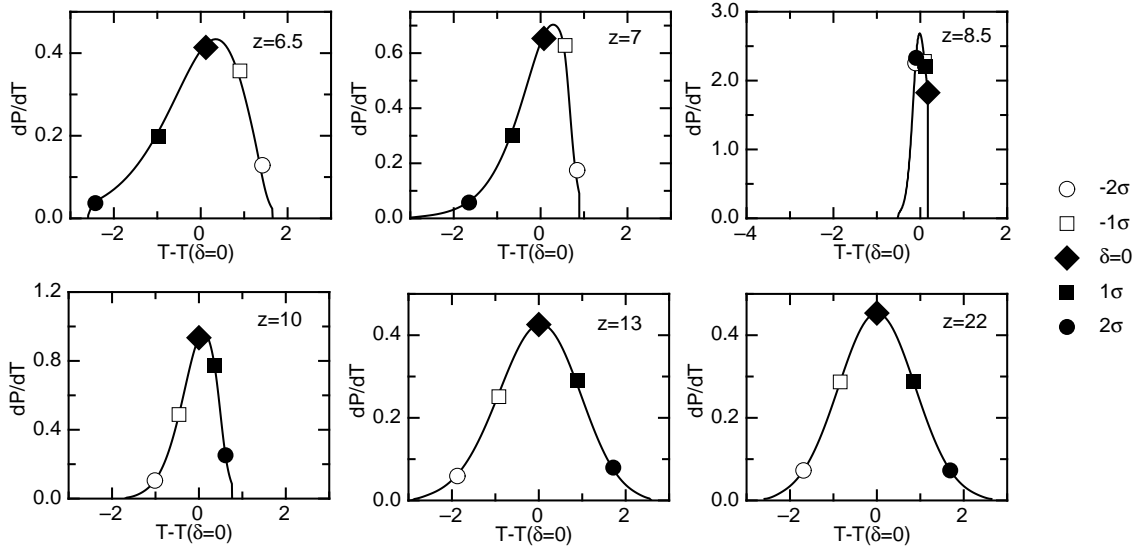} 
\caption{The probability distribution of brightness temperature $dP/dT$. The horizontal axis shows the difference in brightness temperature between over or under-dense regions, and the IGM at mean density [$T-T(\delta=0)$]. Distributions are shown assuming a clumping factor of $C=10$ and an angular scale of 10$^\prime$, at redshifts of $z=6.5$, $z=7$, $z=8.5$, $z=10$, $z=13$ and $z=22$. The large symbols show the value of $T-T(\delta=0)$ corresponding to regions that are $-2\sigma$ (filled circle), $-1\sigma$ (filled square), $0\sigma$ (filled diamond), $1\sigma$ (empty square) and $2\sigma$ (empty circle) fluctuations in the density field and correspond to lines in Figure \ref{fig1}. }
\label{fig2}
\end{figure*}

\begin{figure*}
\includegraphics[width=11.cm]{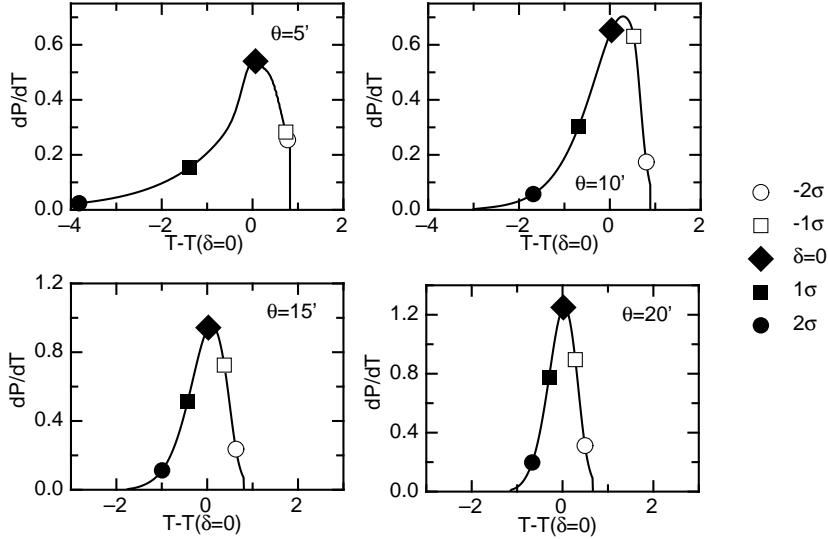} 
\caption{The probability distribution of brightness temperature $dP/dT$. The horizontal axis shows the difference in brightness temperature between over or under-dense regions, and the IGM at mean density [$T-T(\delta=0)$]. Distributions are shown assuming a clumping factor of $C=10$ and a redshift of $z=7$, at angular scales of 5$^\prime$, 10$^\prime$, 15$^\prime$ and 20$^\prime$. The large symbols show the value of $T-T(\delta=0)$ corresponding to regions that are $-2\sigma$, $-1\sigma$, $0\sigma$, $1\sigma$ and $2\sigma$ fluctuations in the density field respectively. }
\label{fig3}
\end{figure*}

Figure~\ref{fig1} also shows that reionisation skews the brightness
temperature towards low values, owing to the dominant effect of galaxy
bias. A point of interest here is that once reionisation begins, this
results in the brightness temperature of both over- and under-dense
regions being lower than the brightness temperature of the IGM at mean
density (we note that the mean is not found to depend on angular scale
as required for consistency). Since the underlying probability
distribution of over-densities $dP/d\delta$ is known (a Gaussian of
variance $\sigma D$), we may compute the observed probability
distribution for $T$,
\begin{equation}
\frac{dP}{dT}(\theta) \propto \frac{dP}{d\delta}\left|\frac{\partial\delta}{\partial T}\right|.
\end{equation}
This distribution corresponds to fluctuations in redshifted 21cm
emission on angular scales $\theta$. Example distributions are shown in
Figures~\ref{fig2} and \ref{fig3}. 

Firstly we discuss the evolution of the distribution with redshift at
fixed angular scale. In Figure~\ref{fig2}, distributions are shown
assuming a clumping factor of $C=10$ and an angular scale of
10$^\prime$, at redshifts of $z=6.5$, $z=7$, $z=8.5$, $z=10$, $z=13$
and $z=22$. The large symbols in this figure show the value of
$T-T(\delta=0)$ corresponding to regions that are $-2\sigma$,
$-1\sigma$, $0\sigma$, $1\sigma$ and $2\sigma$ fluctuations in the
smoothed density field respectively.  The lower right hand panel
clearly shows the Gaussian distribution at high redshift, with evenly
spaced values of $T$ corresponding to the $\pm1\sigma$ and
$\pm2\sigma$ points. At slightly lower redshift, the lower central
panel shows that the Gaussian has become a little wider due to linear
growth, with the high $T$ tail slightly truncated (due to the onset of
skewness), as may be seen by comparing the relative likelihoods for
temperature corresponding to the positive and negative over-density
fluctuations. At a redshift of $\sim10$ in this example (lower left
panel), one gets a narrower and very skewed distribution. At this
point the positive over-densities still correspond to larger than
average brightness temperature, but the difference in brightness
temperature between regions of mean density and $+2\sigma$ fluctuations
is small. The upper right panel shows the distribution at $z=8.5$ (the
minimum). At this redshift, the distribution is both very asymmetric
and narrow, but in addition, the monotonic dependence of brightness
temperature with over-density is broken. Indeed, the mean IGM density
corresponds to the largest brightness temperatures, while the
brightness temperatures of $+1\sigma$ and $-1\sigma$, and of
$+2\sigma$ and $-2\sigma$ density fluctuations are almost the same. At lower redshifts
(upper-central and upper-left panels) the width and skewness of the
distributions both increase, and a monotonic relation between the
over-density and brightness temperature is restored, though the order
is reversed relative to high redshift.

In Figure~\ref{fig3} distributions are shown assuming a clumping
factor of $C=10$ and a redshift of $z=7$, at angular scales of
5$^\prime$, 10$^\prime$, 15$^\prime$ and 20$^\prime$ and $z=7$.  The
distribution is valid on angular scales sufficiently large
that no region on that scale is fully ionised. On small enough scales
the distribution must be described using a model that calculates the size
distribution of individual HII regions (e.g. Furlanetto et
al.~2004). However as may be seen from Figure~\ref{fig3}, our model
describes the distribution $dP/dT$ for $\theta\ga5^\prime$ at $z=7$,
and for all angles of interest to up-coming 21cm experiments at higher
redshifts. At large angular scales (lower-right panel) the
distribution is narrow and close to Gaussian. By comparing the
brightness temperatures corresponding to the $\pm1\sigma$ and
$\pm2\sigma$ regions of IGM with the redshift dependent distributions
in Figure~\ref{fig2} we see that even though the distribution is still
close to Gaussian it already includes the effects of galaxy bias,
since the brighter than average regions are formed by under-dense
regions of IGM. As the angular scale is decreased the width of the
distributions increases, and the skewness towards small $T$ increases.

Figures~\ref{fig2} and \ref{fig3} explicitly demonstrate the
introduction of skewness into the distribution of $T(\theta)$. The
skewness of 21cm intensity maps, and its dependence on model
parameters will be discussed in more detail in the following section.

\section{the variance and skewness of 21cm maps}
\label{moment}

\begin{figure*}
\includegraphics[width=15cm]{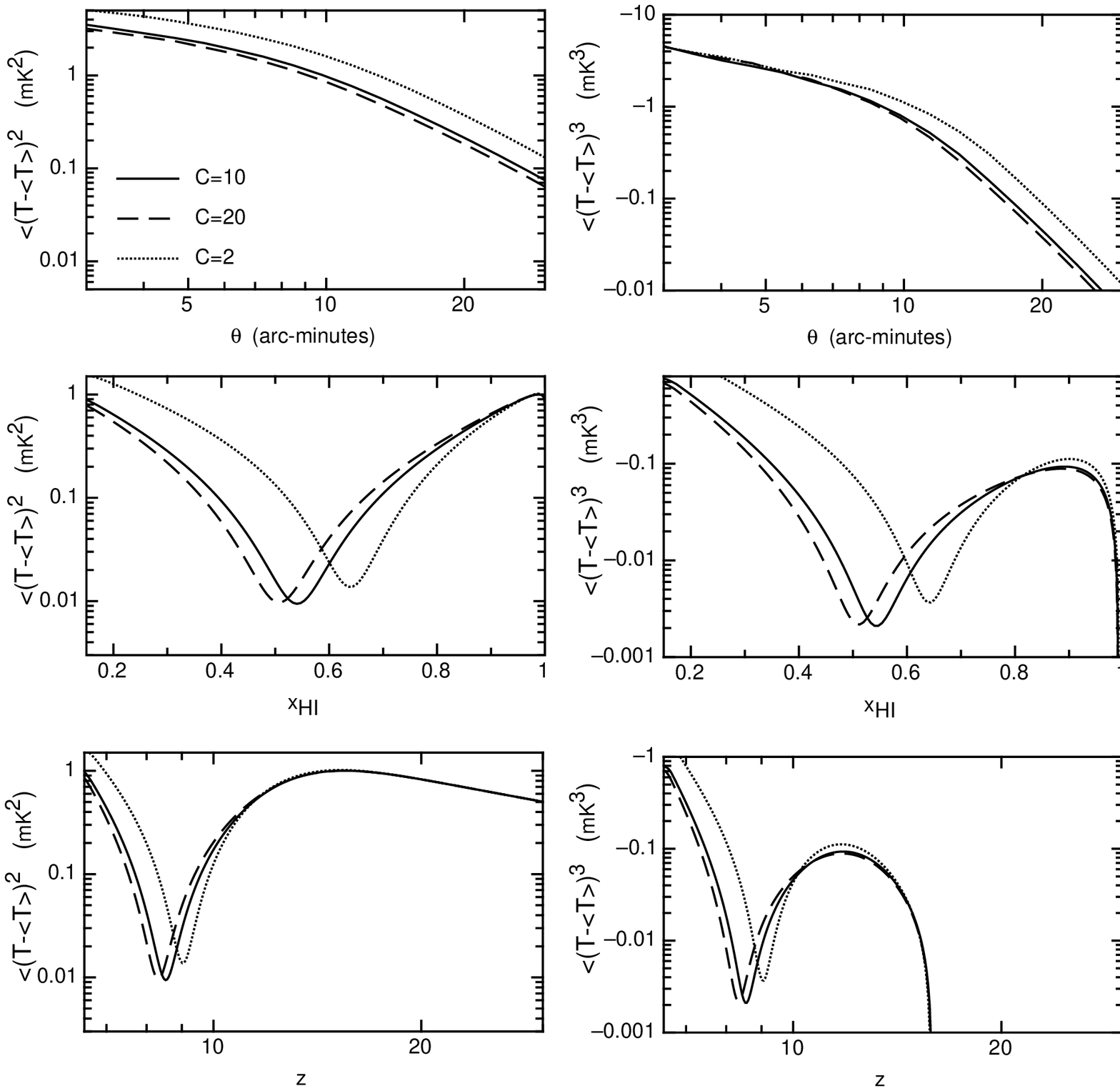} 
\caption{ \textit{Left:} Examples of auto-correlation functions, assuming clumping factors of $C=10$, $C=20$ and $C=2$. The auto-correlation is plotted as a function of $\theta$ (at $z=6.5$, \textit{upper panel}), as a function of mean IGM neutral fraction $x_{\rm HI}$ (at $\theta=10^\prime$, \textit{central panel}), and as a function of redshift (at $\theta=10^\prime$, \textit{lower panel}). \textit{Right:} Corresponding examples of skewness functions.}
\label{fig4}
\end{figure*} 

\begin{figure*}
\includegraphics[width=15cm]{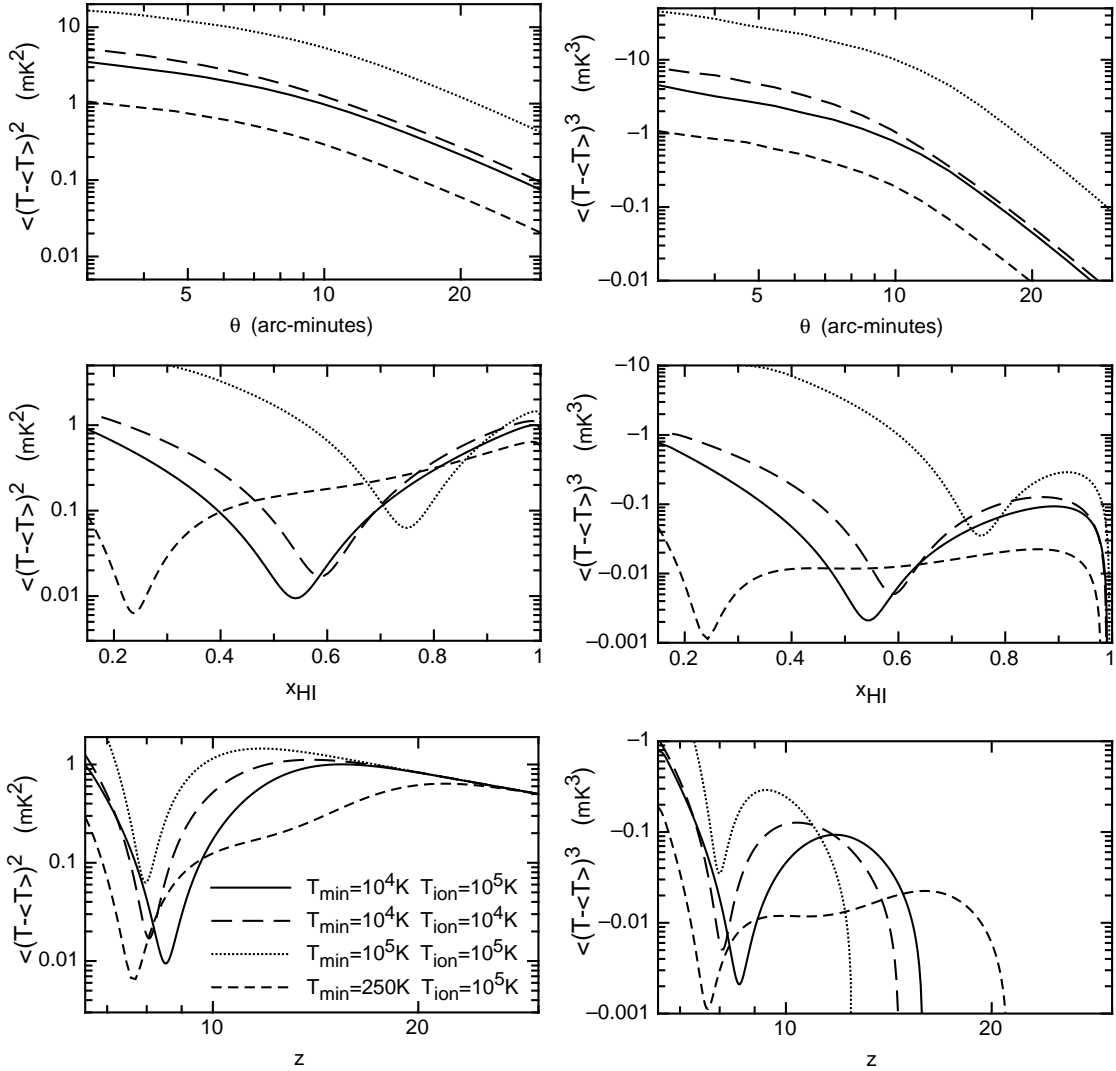} 
\caption{ \textit{Left:} Examples of auto-correlation functions, assuming sets of $(T_{\rm min},T_{\rm ion})=(10^4\mbox{K},10^5\mbox{K})$, $(10^4\mbox{K},10^4\mbox{K})$, $(10^5\mbox{K},10^5\mbox{K})$ and $(250\mbox{K},10^5\mbox{K})$. The clumping factor was chosen to be $C=10$. The auto-correlation is plotted as a function of $\theta$ (at $z=6.5$, \textit{upper panel}), as a function of mean IGM neutral fraction $x_{\rm HI}$ (at $\theta=10^\prime$, \textit{central panel}), and as a function of redshift (at $\theta=10^\prime$, \textit{lower panel}). \textit{Right:} Corresponding examples of skewness functions.}
\label{fig5}
\end{figure*} 

\begin{figure*}
\includegraphics[width=15cm]{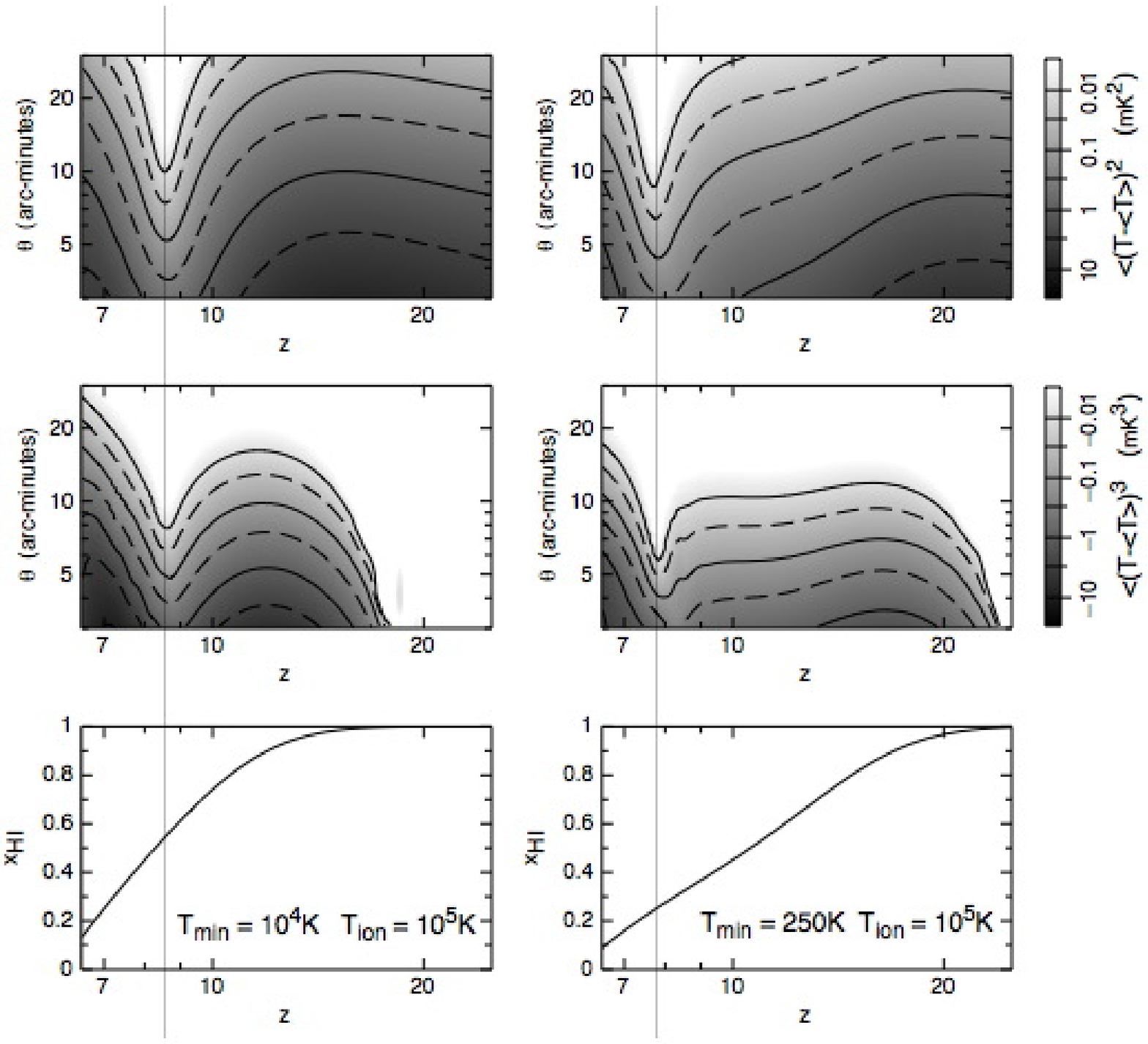} 
\caption{ Contour plots showing the evolution of auto-correlation (\textit{upper panels}) and skewness (\textit{central panels}) functions with redshift and angle, assuming a clumping factor of $C=10$.  Two models for the minimum mass of galaxy formation are shown, $(T_{\rm min},T_{\rm
ion})=(10^4\mbox{K},10^5\mbox{K})$ [left panels] and $(T_{\rm min},T_{\rm
ion})=(250\mbox{K},10^5\mbox{K})$ [right panels]. The keys to the right of the panels show the numerical values assigned to the various contours. The \textit{lower panels} show the evolution of average neutral fraction for these models.}
\label{fig6}
\end{figure*}

In this section we calculate the second and third moments of the
probability distributions of $T$ in redshifted 21cm intensity maps, as
functions of angular scale and redshift. Given the structure of our
semi-analytic model for reionisation, it is natural to discuss the
moments of the real space intensity fluctuations smoothed within
top-hat window functions of angular radius $\theta$.

The second moment of these distributions corresponds to the auto-correlation function of brightness temperature
smoothed on an angular radius $\theta$:
\begin{eqnarray}
\nonumber
\langle \left(T-\langle T\rangle\right)^{2}\rangle &=&\\
 &&\hspace{-15mm}\left[\frac{1}{\sqrt{2\pi}\sigma(R)}\int d\delta \left(T(\delta,R)-\langle T\rangle\right)^2e^{-\frac{\delta^2}{2\sigma(R)^2}} \right],
\end{eqnarray}
where 
\begin{equation}
\langle T\rangle = \frac{1}{\sqrt{2\pi}\sigma(R)}\int d\delta~ T(\delta,R)e^{-\frac{\delta^2}{2\sigma(R)^2}} ,
\end{equation}
and $\sigma(R)$ is the variance of the density field (at redshift $z$) smoothed on a scale
$R$. 

A measure of the departure of the statistics from those of a Gaussian
random field is provided by the third moment of the distribution
$dP/dT$. We refer to the third moment as the skewness function
 of brightness temperature smoothed with top-hat
windows of angular radius $\theta$.
\begin{eqnarray}
\nonumber
\langle \left(T-\langle T\rangle\right)^{3}\rangle&=&\\
 &&\hspace{-15mm} \left[\frac{1}{\sqrt{2\pi}\sigma(R)}\int d\delta
 \left(T(\delta,R)-\langle
 T\rangle\right)^3e^{-\frac{\delta^2}{2\sigma(R)^2}}
 \right].
\end{eqnarray}

Examples of auto-correlation and skewness functions are shown in the
left and right hand panels of Figure~\ref{fig4} assuming clumping
factors of $C=10$, $C=20$ and $C=2$. The auto-correlation function is
plotted as a function of $\theta$ (at $z=6.5$, upper panel), as a
function of mean IGM neutral fraction $x_{\rm HI}$ (at
$\theta=10^\prime$, central panel), and as a function of redshift (at
$\theta=10^\prime$, lower panel). In the left and right hand panels of
Figure~\ref{fig5} we show examples of auto-correlation and skewness
functions for four commonly discussed scenarios of galaxy formation in
a neutral and ionised IGM. The four cases presented in Figure~\ref{fig5}
show results for the following combinations of $(T_{\rm min},T_{\rm ion})$: $(10^4\mbox{K},10^5\mbox{K})$
corresponding to our fiducial model with star-formation down to the
hydrogen cooling threshold in neutral regions plus radiative
feedback; $(10^4\mbox{K},10^4\mbox{K})$ corresponding to a model with
no radiative feedback; and $(10^5\mbox{K},10^5\mbox{K})$ which could
correspond to a model where feedback within the galaxy (e.g. from
supernovae Dekel \& Woo~2003) limits the star formation efficiency in
low mass galaxies; and $(250\mbox{K},10^5\mbox{K})$ corresponding to a
model where molecular hydrogen cooling allows star formation to occur
in mini-halos below the hydrogen cooling threshold. The clumping
factor was assumed to be $C=10$ in each case. In all examples presented in Figures~\ref{fig4}-\ref{fig5} we have
set the efficiency of the ionising sources so as to achieve reionisation in
the mean IGM at $z=6$.

We find that the amplitude of the auto-correlation function decreases
with angular scale as expected. The amplitude of the skewness also
decreases with angular scale. However interestingly, the amplitudes of
the auto-correlation and skewness functions do not vary monotonically
with redshift or average neutral fraction. In particular, both the
variance and skewness of the 21cm intensity field have minima during
the reionisation history.  At high redshift the skewness goes to zero,
since the 21cm map prior to galaxy formation must follow a Gaussian
distribution corresponding to the primordial density fluctuations.

This unintuitive behavior was alluded to in Figure~\ref{fig1} and can
be understood as follows. Early in the reionisation history, before
galaxies have had much influence on the ionisation state of the IGM,
the variance is governed by the primordial power-spectrum of
fluctuations in the density field, and the skewness is zero. During
this early time, the over-dense regions are brighter due to the higher
density of neutral IGM there. As galaxies modify the ionisation state
of the IGM, the over-dense regions get ionised first, breaking the
Gaussianity of the 21cm map. Therefore, once reionisation is underway,
the over-dense regions of IGM become fainter in 21cm emission relative
to under-dense regions. As was shown in Figure~\ref{fig1}, this
results in a transition between bright patches of 21cm emission being
observed in over-dense regions, too bright patches being observed in
under-dense regions. This transition yields an epoch of nearly uniform
brightness temperature of the IGM when smoothed at a certain scale, and hence minima in the
auto-correlation and skewness functions. At later times, both the
variance and the skewness increase as reionisation is completed.

To illustrate this complex behavior in more detail, we plot contours
of the auto-correlation and skewness in Figure~\ref{fig6} as functions
of redshift and angular scale (upper and central panels). For
comparison we also plot the evolution of the average neutral fraction
with redshift (lower panels). Two cases are shown, corresponding to
the two models in Figure~\ref{fig5} that include radiative feedback,
i.e. $(T_{\rm min},T_{\rm ion})=(10^4\mbox{K},10^5\mbox{K})$ [left
panels] and $(T_{\rm min},T_{\rm ion})=(250\mbox{K},10^5\mbox{K})$
[right panels]. Comparison of the upper and central panels with the
lower panels in Figure~\ref{fig6} shows that minima in both the
auto-correlation and skewness occur when the ionised fraction is
around $Q\sim0.5-0.7$, near a redshift of $z\sim8-9$. The minimum
occurs at higher redshifts for fluctuations measured at smaller angular scales.
Comparison of the central and lower panels of Figure~\ref{fig6} show
that skewness appears in the 21cm map at the redshift where the
reionisation of the IGM begins (at $z\sim20$ in this model). However
the upper panels show non-zero fluctuations in 21cm emission at
earlier times, which are driven by the density fluctuations in the
IGM. Note that the two models yield identical fluctuation statistics at $z\sim20$. Finally, Figure~\ref{fig6} shows that the skewness becomes larger
in comparison to the auto-correlation amplitude as reionisation
proceeds.

In Figures~\ref{fig4} and \ref{fig5} we demonstrated the dependence of
the auto-correlation and skewness functions on model parameters as a
function of angle or redshift. Figure~\ref{fig6} demonstrates this
dependence more-fully. While the general trends and overall shape are
similar in both cases, there are substantial quantitative
differences. These plots demonstrate the rich detail on the
reionisation history that can be found by considering both the
auto-correlation and skewness functions and their evolution.  In
particular, the amplitudes of the auto-correlation and skewness
functions, as well as the epoch and depth of their minima in redshift
are sensitive to the details of the reionisation history. The shape of
the angular power-spectrum has been shown to encode information on the
topology of the reionisation process. Our results show that the
evolution of the fluctuation statistics in 21cm intensity maps will
provide complementary information on the astrophysics of the
reionisation epoch. This is the topic of the next section.

\subsection{Variation of auto-correlation and skewness with model
parameters}

The free parameters that govern the details of the reionisation
history described in equation~(\ref{history}) are the clumping factor
$C$ and the minimum masses for galaxy formation in a neutral ($M_{\rm
min}$) and reionised IGM ($M_{\rm ion}$). These parameters encapsulate
the astrophysics of the IGM, and of the connection between the IGM and
ionising sources respectively. In this section we discuss the
variation of the auto-correlation and skewness functions with these
parameters.

In Figure~\ref{fig4} we demonstrated the variation of the angular
auto-correlation and skewness functions with clumping of the IGM. A
larger clumping factor results in an increased recombination rate in
over-dense regions. At low redshift, where over-dense regions have lower
than average 21cm intensity, this increased recombination rate reduces
the effect of galaxy bias on the variation of 21cm intensity with
over-density by delaying the reionisation of over-dense regions. Thus
the observational effect of a large clumping factor in the IGM is to
reduce both the variance and the skewness of the distribution of 21cm
intensity fluctuations. Conversely at higher redshifts (where
over-dense regions have higher than average 21cm intensity), the
increased recombination rate results in an increased amplitude for the
auto-correlation and skewness functions.

In Figure~\ref{fig5} we demonstrate the dependence of the angular
auto-correlation and skewness functions on the minimum mass for galaxy
formation in a neutral and ionised IGM. The four cases shown
correspond to commonly discussed possibilities for the star formation
history at high redshift. Our fiducial model includes radiation
feedback by increasing the virial temperature corresponding to the
minimum mass from $T_{\rm min}$ to $T_{\rm ion}$ following
reionisation of a region. As a result, once part of a region becomes
ionised, the photon production there is suppressed, slowing down the
subsequent ionisation rate relative to more neutral regions. Comparison
of the fiducial model having $(T_{\rm min},T_{\rm
ion})=(10^4\mbox{K},10^5\mbox{K})$ with the case of no feedback
$(T_{\rm min},T_{\rm ion})=(10^4\mbox{K},10^4\mbox{K})$ illustrates
the suppression of both the amplitude of the auto-correlation and
skewness functions in the presence of radiative feedback. The redshift
where the minimum values of auto-correlation and skewness are reached
is raised by the inclusion of feedback (given a common redshift for
the completion of reionisation).  The signature of radiative feedback
is seen in the evolution of auto-correlation and skewness amplitude
with neutral fraction. While there is substantial difference in the
predicted amplitudes at neutral fractions below 0.5, the two cases are
very similar before reionisation reaches a significant level. Thus the
signature of radiative feedback is to suppress the amplitude of
fluctuations only late in the reionisation epoch.

In a model where the minimum mass for galaxy formation is $10^5$K in
all regions ($T_{\rm min}=T_{\rm ion}=10^5$K), the amplitudes of both
the auto-correlation and skewness are substantially increased (relative
to the fiducial model) at all redshifts following the onset of
reionisation. Similarly, a model with $T_{\rm min}=250$K and $T_{\rm
ion}=10^5$K predicts amplitudes of both the auto-correlation and
skewness that are substantially smaller than the fiducial
model. Indeed the variation in amplitude of the auto-correlation at fixed
redshift reaches an order of magnitude over the astrophysically
plausible range of minimum masses. This prediction of the variation of
fluctuation amplitude with the mass of ionising sources is consistent
with the numerical simulations of Lidz et al.~(2006). The variation is
due too the larger (smaller) bias of the ionising sources, which leads
to a larger (smaller) variation in ionisation state between over-dense
and under-dense regions. This larger (smaller) variation in turn leads
to a larger (smaller) amplitudes for both the auto-correlation and
skewness functions.

All models in Figure~\ref{fig5} show minima of the variance and
skewness at redshifts of $z\sim8-9$. The duration and depth of these
minima are sensitive to the bias of the ionising sources. In
particular, the ratio of the auto-correlation amplitude at high redshift to the
auto-correlation amplitude at the minimum varies between a factor of 10 and a
factor of 100 across the models considered. Similar results are found
for the skewness. The redshift of the minimum is also sensitive to the
bias of the ionising sources, though the dependence is not
monotonic. On the other hand there is a monotonic relation between the
neutral fraction at the redshift of the minimum and the bias, with
more massive ionising sources generating a minimum at a smaller
ionised fraction. Models with low clumping and models with large
source mass both increase the amplitude of the auto-correlation and
skewness. This degeneracy between models may be broken by looking at
the location of the minima in variance and skewness as functions of
redshift and neutral fraction. Assuming a fixed epoch of reionisation,
both large clumping factors and massive ionising sources delay the
minima in time. Conversely, a large clumping factor results in the
minima occurring at a smaller average neutral fraction, while massive
ionising sources lead to the minima occurring at a higher neutral
fraction. However in practice the neutral fraction will be very
hard to measure due to instrumental and foreground effects, which may
limit the utility of this effect.

In summary the evolution of the variance and skewness in 21cm
intensity maps with redshift and mean neutral fraction will probe
physical quantities like the clumping of the IGM and the mass of
ionising sources.  In particular, reionisation scenarios with smaller
values of clumping factor, or more massive (and therefore more biased)
ionising sources result in 21cm intensity maps with larger values of
variance and skewness. Indeed, the variation in amplitude of the
auto-correlation at fixed redshift reaches an order of magnitude over
the astrophysically plausible range of minimum masses. In addition,
the relative locations of minima in variance and skewness when plotted
as functions of redshift and neutral fraction can be used to
distinguish between amplitudes that are increased due to large values
of ionising source mass, or small values of clumping factor. While
more massive sources increase the amplitude of fluctuations at all
epochs, the signature of radiative feedback is the suppression of
fluctuations at epochs following, but not prior to the substantial
reionisation of the IGM. We therefore conclude that the detection of
fluctuations in 21cm intensity should unambiguously determine the mass
of ionising sources.

\section{Measurement of Skewness in 21cm maps}
\label{measure}

In the previous sections we have demonstrated that galaxy bias should
lead to skewness in the intensity distribution derived from redshifted
21cm maps of the reionisation epoch. In this section we discuss the
prospects for measuring that skewness using first generation
instruments.

\begin{figure*}
\includegraphics[width=15cm]{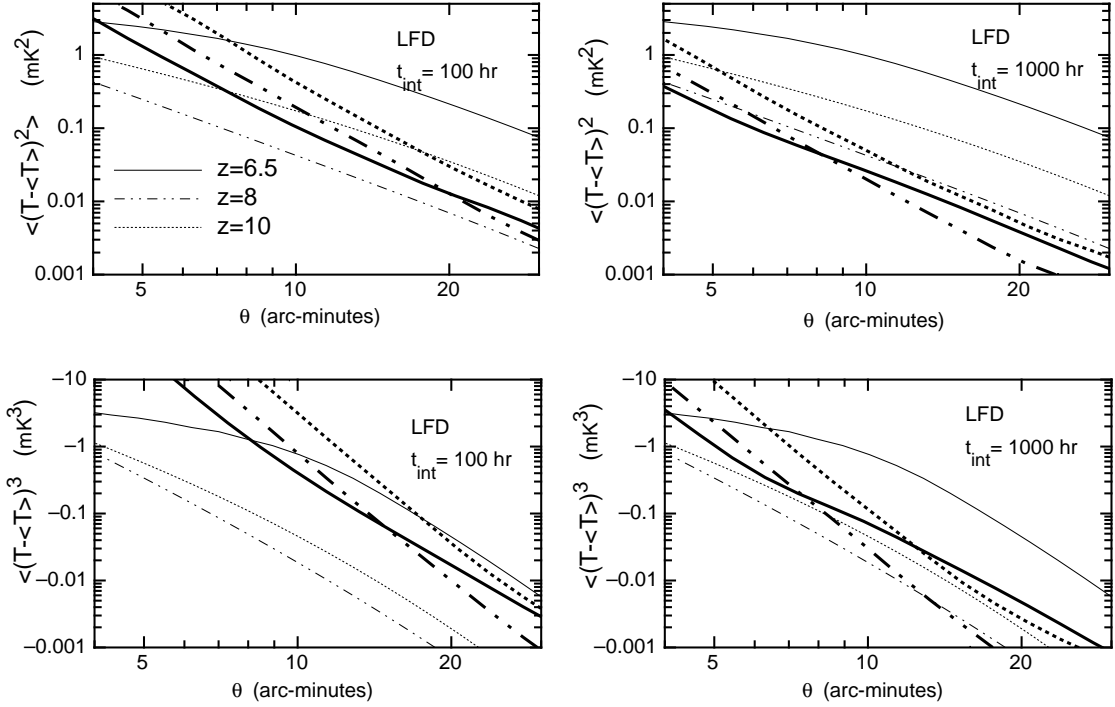} 
\caption{\textit{Upper panels:} Comparisons of signal and noise for the
auto-correlation as a function of angle assuming
$C=10$, and at $z=6.5$, $z=8$ and $z=10.0$. The thin and thick lines correspond to signal and noise respectively. \textit{Lower panels:} The corresponding comparisons for skewness. In each case the \textit{left} and \textit{right} hand panels correspond to integration times of 100 hours and 1000 hours respectively.}
\label{fig7}
\end{figure*} 

\begin{figure*}
\includegraphics[width=15cm]{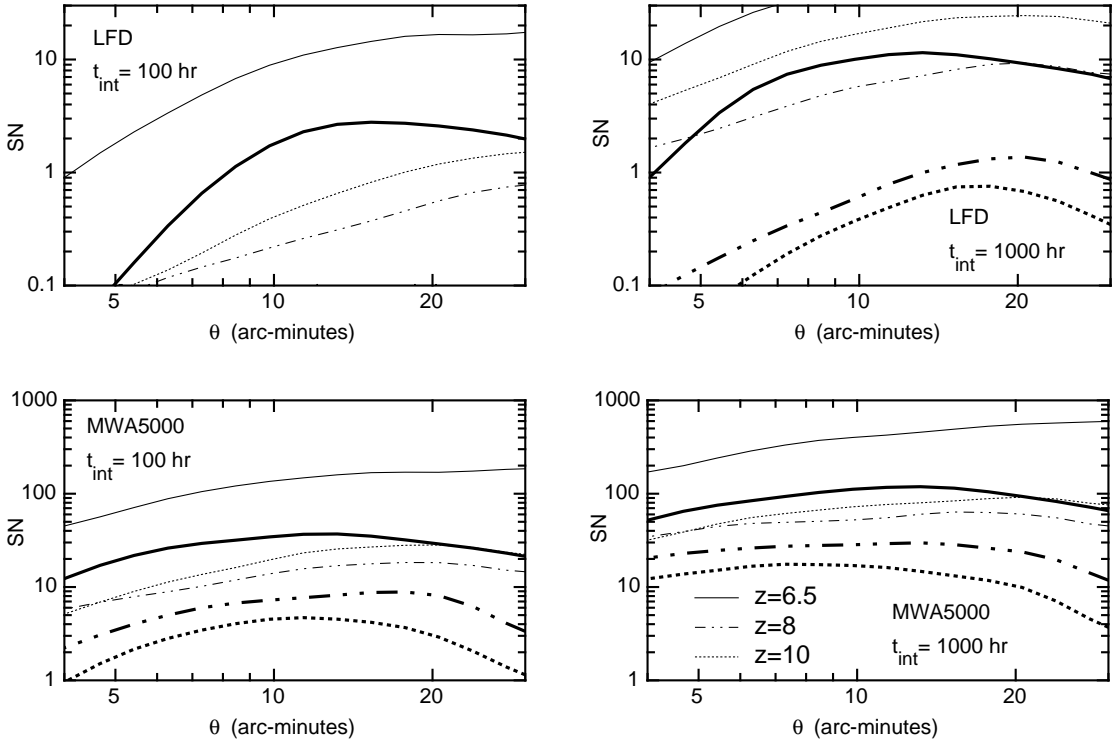} 
\caption{The signal-to-noise ratios for the auto-correlation (thin lines) and skewness (thick lines) functions. In each case the \textit{left} and \textit{right} hand panels correspond to integration times of 100 hours and 1000 hours respectively. The \textit{upper} and \textit{lower} rows show results for the LFD and MWA5000 respectively.}
\label{fig8}
\end{figure*} 

We first discuss the response of a phased array to the brightness
temperature contrast of the IGM. Assuming that calibration can be performed
ideally, and that foreground subtraction is perfect, the root-mean-square
fluctuations in brightness temperature are given by the radiometer equation
\begin{equation}
\Delta T = \frac{\epsilon\lambda^2T_{\rm sys}}{A_{\rm
tot}\Omega_{\rm b}\sqrt{t_{\rm int}\Delta\nu}},
\end{equation}
where $\lambda$ is the wavelength, $T_{\rm sys}$ is the system
temperature, $A_{\rm tot}$ the collecting area, $\Omega_{\rm b}$ the
effective solid angle of the synthesised beam in radians, $t_{\rm
int}$ is the integration time, $\Delta\nu$ is the size of the
frequency bin, and $\epsilon$ is a constant that describes the overall
efficiency of the telescope. We optimistically adopt $\epsilon=1$ in this
paper. 

As a concrete example we consider the case of the Mileura Wide Field Array--Low Frequency Demonstrator (hereafter LFD\footnote{see
http://www.haystack.mit.edu/ast/arrays/mwa/index.html}), though
the measurement described could also be made using other instruments
such as LOFAR\footnote{http://www.lofar.org/}. In units relevant for the LFD, and at $\nu=200$MHz,
we find (Wyithe, Loeb \& Barnes~2005)
\begin{eqnarray}
\label{noiseeqn}
\nonumber
\Delta T &=& 7.5\left(\frac{T_{\rm sys}}{250\mbox{K}}\right)\mbox{mK} \left(\frac{1.97}{C_{\rm beam}}\right)\left(\frac{A_{\rm tot}}{A_{\rm LFD}}\right)^{-1}\\
&\times&\left(\frac{\Delta\nu}{1\mbox{MHz}}\right)^{-1/2}\left(\frac{t_{\rm int}}{100\mbox{hr}}\right)^{-1/2}\left(\frac{\theta_{\rm beam}}{5^\prime}\right)^{-2}.
\end{eqnarray}
Here $A_{\rm LFD}$ is the collecting area of a phased array consisting of
500 tiles each with 16 cross-dipoles [the effective collecting area of an
LFD tile with $4\times4$ cross-dipole array with 1.07m spacing is
$\sim17-19$m$^2$ between 100 and 200MHz (B. Correy, private
communication; Bowman, Morales \& Hewitt 2005)]. The system temperature will be dominated by the
sky and in the frequency range of interest has a value $T_{\rm sys}\sim280[(1+z)/7.5]^{2.3}$K.  $\Delta\nu$ is the frequency
range over which the signal is smoothed and $\theta_{\rm beam}$ is the size
of the synthesised beam. The value of $\theta_{\rm beam}$ can be regarded
as the radius of a hypothetical top-hat beam, or as the variance of a
hypothetical Gaussian beam. The corresponding values of the constant
$C_{\rm beam}$ are 1 and 1.97 respectively. 

Equation~(\ref{noiseeqn}) gives the uncertainty $\Delta T$ in the 21cm
signal measured from a cylinder of IGM, with radius $\theta$ and depth
$\Delta\nu$. For comparison with our calculated auto-correlation and skewness, we
modify this in order to estimate the noise in the signal originating from a spherical
region of radius $R$, so that $\theta = R/d_{A}$ and $\Delta\nu/\nu=
2R(cdt/dz)^{-1}(1+z)^{-1}$ where $\nu$ is the redshifted 21cm
frequency. We find that consideration of spherical regions increases
the noise relative to equation~(\ref{noiseeqn}) by a factor of
$\sqrt{15/8}$.

In any one synthesised beam, the measurement noise will be far in
excess of the 21cm signal. However as has been discussed
previously (Bowman, Morales \& Hewitt 2005), the determination of the power-spectrum may still
be made since it can be constructed from a large number of independent
measurements (each of signal-to-noise below unity). In practice one could determine the auto-correlation function (or power-spectrum) by dividing the data stream in half, and performing a cross-correlation (Max Tegmark, private communication). Hence defining $T$ as the measurement signal, and $\Delta T$ as the thermal noise, we have
\begin{eqnarray}
\nonumber
\label{ac}
&&\hspace{-3mm}\langle[(T-\langle T\rangle)+\Delta T_1][(T-\langle T\rangle)+\Delta T_1]\rangle\\
\nonumber
&&\hspace{-0mm} =\langle(T-\langle T\rangle)^2\rangle\\
&&\hspace{-0mm} + \langle(T-\langle T\rangle)\Delta T_1\rangle + \langle(T-\langle T\rangle)\Delta T_2\rangle + \langle\Delta T_1\Delta T_2\rangle, 
\end{eqnarray}
where $\Delta T_1$ and $\Delta T_2$ are drawn from Gaussian uncertainty distributions of width $\sqrt{2}\Delta T$. Here, with the exception of the first term (which is the variance), the cross-correlation results in terms that average to zero in the limit of large numbers of measurements. In a finite experiment, the sum of these terms is the measurement noise. 

To estimate the noise in the measurement, we have made many Monte-Carlo realisations of
the observed distribution. For each measurement we obtain $N$ values
of observed brightness temperature by adding telescope noise to the
intrinsic $T$ drawn at random from the theoretical distribution.  The
error for each measurement can then be calculated from the noise terms
in the cross-correlation, and the uncertainty in the measurement estimated as the variance
among the values of error. For a field of angular size $\theta_{\rm
field}$ and a Gaussian synthesised beam of radius $\theta$, the number
of independent pointings is approximated by $N=[\theta_{\rm field}/(2
\theta)]^2$, where the 2 in the denominator reduces the effect of
overlapping synthesised beams. We assume $\theta_{\rm field}=10^o$ for
the LFD.

Comparisons of signal and noise are plotted in the upper panels of
Figure~\ref{fig7}. In both the left and right panels, the
auto-correlation is plotted as a function of angle assuming
$C=10$, and at $z=6.5$, $z=8$ and $z=10.0$ (thin lines). For
comparison we plot the estimate of noise (thick lines), assuming
integrations of 100 hours (left panel) and 1000 hours (right
panel). We find that the auto-correlation function can be detected
at redshifts below the minimum in 100hr.  In 1000 hours, higher
redshift fluctuations can also be measured.

Our signal to noise estimate is approximate, and does not account for
several important issues. For example, the LFD will have a very
compact antenna arrangement. As a result the noise is not independent
in all synthesised beams. This is because short baselines only give
independent information on length scales associated with their
separation. Therefore with a compact arrangement, it is only at very
large scales that the pixels become independent. Moreover, we have
only considered a single frequency slice, whereas fluctuations can be
measured in many slices at slightly different redshifts. This latter
effect will tend to increase the sensitivity achievable relative to
our estimate, while the former will tend to reduce it.

In analogy with the above procedure one could determine the skewness function by dividing the data stream in three, and performing a cross-correlation. Hence
\begin{eqnarray}
\label{skew}
\nonumber
&&\hspace{-7mm}\langle[(T-\langle T\rangle)+\Delta T_1][(T-\langle T\rangle)+\Delta T_2][(T-\langle T\rangle)+\Delta T_3]\rangle \\
\nonumber
&&\hspace{-3mm} = \langle(T-\langle T\rangle)^3\rangle\\
\nonumber
&&\hspace{-3mm} + \langle(T-\langle T\rangle)^2\Delta T_1\rangle + \langle(T-\langle T\rangle)^2\Delta T_2\rangle + \langle(T-\langle T\rangle)^2\Delta T_3\rangle\\
\nonumber
&&\hspace{-3mm} + \langle(T-\langle T\rangle)\Delta T_1\Delta T_2\rangle + \langle(T-\langle T\rangle)\Delta T_1\Delta T_3\rangle\\
&&\hspace{-3mm} + \langle(T-\langle T\rangle)\Delta T_2\Delta T_3\rangle + \langle\Delta T_1\Delta T_2\Delta T_3\rangle, 
\end{eqnarray}
where $\Delta T_1$, $\Delta T_2$ and $\Delta T_3$ are drawn from
Gaussian uncertainty distributions of width $\sqrt{3}\Delta T$. As
before, with the exception of the first term (which is the skewness),
the cross-correlation results in terms that average to zero in the
limit of large numbers of measurements.  In analogy with the above, we
have made many Monte-Carlo measurements of the error from the noise
terms in the cross-correlation, and estimated the uncertainty in the
measurement of skewness form the variance among these Monte-Carlo
errors.

Comparisons of signal and noise for the skewness are plotted in the lower panels of
Figure~\ref{fig7}. In both the left and right panels, the skewness is
plotted as a function of angle assuming $C=10$, and at $z=6.5$,
$z=7.0$, $z=8$ and $z=10.0$ (thin lines). These curves correspond to
the auto-correlation curves described above. For comparison we again
plot the estimate of noise (thick lines), assuming integrations of 100
hours (left panel) and 1000 hours (right panel). In 100 hours, the
skewness could only be (marginally) detected at lower redshifts, and
at angles near $\theta\sim12^\prime$. However in 1000 hours, the
skewness can be detected over a range of angles at redshifts below the
minimum.

Estimates of the signal-to-noise ($SN$) for measurements of the
auto-correlation and skewness were obtained by taking the ratio of the
signal to the root-mean-square of the sum of the noise terms in
equations~(\ref{ac}-\ref{skew}).  The upper panels of
Figure~\ref{fig8} show estimates of $SN$ as a function of $\theta$ for
integrations of 100 hours and 1000 hours using the LFD. The thin lines
show $SN$ for the auto correlation function, while the thick lines
show results for the skewness. In 1000 hours and at low redshift, the
skewness can be detected by the LFD at a signal-to-noise as large as
$SN\sim10$. The $SN$ for detection of skewness is a factor of several
to 10 below that achievable for the auto-correlation function.

The above results suggest that the skewness in redshifted 21cm maps
should be just detectable with the LFD. However a hypothetical follow
up to the LFD would comprise $\sim10$ times the collecting area, with
baselines of $\sim10$s to $100$s of kilometers (we refer to this as
the MWA-5000). The signal-to-noise achievable with the MWA-5000 is plotted as a function of
$\theta$ for integrations of 100 hours and 1000 hours. In 100 hours, an MWA-5000 could detect skewness at high
signal to noise outside the minimum redshift. In 1000 hours an MWA-5000 could
detect skewness at all redshifts, and map out its evolution in
detail. Detection of skewness would therefore be an important
scientific driver for a second generation array.

\section{conclusion}
\label{conclusion}

Using a simple semi-analytic model for density dependent reionisation,
we have quantified the extent to which galaxy bias induces skewness in
the intensity distributions of redshifted 21cm maps. Our model
demonstrates that skewness of the distribution is present on all
scales, and during the entire reionisation era. In particular,
skewness is present on scales where no single region is fully
reionised. These results are consistent with the detailed numerical
modeling previously presented by Lidz et al.~(2006). The skewness
(like the variance) decreases with increasing angular scale, but is
not monotonic with redshift or neutral fraction, even in a simple
reionisation scenario.  A feature of our density dependent
reionisation model is that it predicts a minimum in both the variance
and skewness at the time when the ionisation fraction is of order
50\%. This minimum is a generic feature of models that include galaxy
bias, and is due to the transition between the high redshift era when
the 21cm fluctuations are generated by the density field, and the
lower redshift era when fluctuations are dominated by biased
reionisation of the IGM.

The large scale skewness described in this paper is related to, but is
different from from the excess power (or bump) induced in the
power-spectrum at small scales following the appearance of bubbles
(Furlanetto et al.~2004). We are unable to describe the statistics of
small scale fluctuations which require full numerical modeling (see
e.g. Lidz et al.~2006). Skewness appears in regions of low to moderate
average ionisation, and is a result of galaxy bias arising from galaxy
formation within large scale over-densities. We note that if the
ionisation fraction was independent of density, or even if it was a
linear function of over-density, then there would be no
skewness. Indeed the relation between ionisation fraction and density
would be linear in the absence of galaxy bias (and of
recombinations). However galaxy bias ensures a non-linear relation,
and as a result, skewness of the 21cm intensity maps will be present
throughout reionisation, and on all scales.

We have discussed the evolution of the amplitudes of the
auto-correlation and skewness of 21cm intensity maps, as well as their
evolution with redshift and mean IGM neutral fraction.
The evolution of the variance and skewness in 21cm intensity maps will
probe quantities like the clumping of the IGM, the mass of ionising
sources and the presence of radiative feedback.  For example, we find
that reionisation scenarios with smaller values of clumping factor, or
more massive (and therefore more biased) ionising sources result in
21cm intensity maps with larger values of variance and
skewness. Indeed the variation in amplitude of the auto-correlation at
fixed redshift reaches an order of magnitude over the astrophysically
plausible range of minimum masses.  Moreover, the relative locations of
minima in variance and skewness when plotted as functions of redshift
and neutral fraction can distinguish between amplitudes that are
increased due to large values of ionising source mass, or small values
of clumping factor. In addition, while more massive sources increase the amplitude
of fluctuations at all epochs, the signature of radiative feedback is
the suppression of fluctuations at epochs following, but not prior to
the substantial reionisation of the IGM. We therefore conclude that
the detection of fluctuations in 21cm intensity should therefore
unambiguously determine the mass of ionising sources.

We have estimated the sensitivity of the Mileura Widefield Array
Low-Frequency Demonstrator to skewness in 21cm intensity maps, and
find that the predicted level should be detectable in long
integrations ($\sim1000$ hours) at redshifts nearing the end of
reionisation.  However an instrument with 10 times the collecting area
of the LFD could detect skewness across a range of redshifts. Skewness in 21cm
maps, and a minimum in the variance during the reionisation era are
generic predictions of models of reionisation that account for biased
galaxy formation. The presence of this minimum in variance, and of
skewness, which will be detectable at a signal-to-noise a few to ten
times smaller than the variance will provide confirmation of the
cosmological nature of any observed angular fluctuations.

{\bf Acknowledgments} The research was supported by the Australian Research
Council (JSBW), and by National Science Foundation grant \#0457585 (MFM).

\newcommand{\noopsort}[1]{}

\label{lastpage}
\end{document}